\documentstyle[12pt]{article}
\begin{document}
\newcommand{\rav}{\:\hbox to 10.7pt {=\hss$\bigcirc$}}
\newcommand{\pls}{\:\hbox to 10.7pt {+\hss$\bigcirc$}}
\title{\large \hspace{10cm} ITEP-29 \\ \hspace{10cm} August 1998 \\
\vspace{1cm}
\LARGE \bf Hydrogen atom in crossed external fields reexemined by
the moment method}
\author {V. A. GANI\thanks{E-mail: gani@vxitep.itep.ru}{\,}
\\
{\it Moscow State Engineering Physics Institute (Technical University),}\\
{\it Kashirskoe shosse, 31, Moscow, 115409, Russia}\\
{\it and}\\
{\it Institute of Theoretical and Experimental Physics, Russia}\\
\\
V. M. WEINBERG \thanks{E-mail: wein@vxitep.itep.ru},
\\
{\it Institute of Theoretical and Experimental Physics,}\\
{\it B.Cheremushkinskaja, 25, Moscow, 117259, Russia}\\
}
\date{}
\maketitle
\vspace{1mm}
\centerline{\bf {Abstract}}
\vspace{3mm}
Recurrence relations of perturbation theory for hydrogen ground state
are obtained. With their aid polarizabilities in constant
perpendicular electric and magnetic fields are computed up to 80th order.
The high orders asymptotic is compared with its quasiclassical estimate.
For the case of arbitrary mutual orientation of external fields a general
sixth order formula is given.

\newpage

\begin{center}
\bf
1. Introduction
\end{center}

   The hydrogen atom in constant homogeneous electric and magnetic
fields still remains to be an object of theoretical investigations.
A good example is the recent work~\cite{klj}, where a recurrent
nonperturbative method is developed for building the exact wave function
of hydrogen atom in magnetic field in the form of convergent double
series. More wide discussion of the problem is contained  in the
review~\cite{l}.

   The famous technical trouble, namely the inability to separate the
variables, only stimulate application of new investigation methods,
including perturbative ones. The moment method~\cite{a}, first used
for perturbative treatment of the anharmonic oscillator, is not
related with variables separation.
The recent application of this method to the Zeeman effect problem~\cite{gkw}
allowed to check the behavior of high orders asymptotic of the perturbation
series. The moment method in the form similar to that used in~\cite{gkw},
was developed independently of Ader's work~\cite{a} by Fernandez and
Castro~\cite{fc}. Then it was applied to hydrogen atom placed in
parallel electric and magnetic fields~\cite{fc2} and 
later the Zeeman effect problem was considered for four sequences of hydrogen 
atom states~\cite{fm}.

     It seems to be even more
important to apply it to hydrogen atom in crossed electric (${\cal
\vec E}$) and magnetic (${\cal \vec H}$) fields because only initial
terms of expansion in powers of ${\cal \vec E}$ and ${\cal \vec H}$
were considered for this case up to now~\cite{dmo,lvhk,s,t,mmo}.
As will be shown here the moment method allows to compute high enough
orders of this expansion.

   The high orders asymptotic can be obtained with the help of the
imaginary time method~\cite{ppt,pp,pkp}. This asymptotic is
determined by the contribution of an extreme subbarrier classical
trajectory into the atom ionization probability~\cite{bw,ps}. A pair
of extreme paths replaces this trajectory at some value of the ratio
of external fields $\gamma={\cal H}/{\cal E}$. The dependence of far
perturbation series terms on $\gamma$ reflects this change of
extreme trajectory and should be especially sharp for perpendicular
external fields. We study here just this case.

\begin{center}
\bf
2. Recurrent evaluation of perturbation series
\end{center}

   Consider the ground state of hydrogen atom, placed in
perpendicular electric and magnetic fields. These fields are supposed
to be constant and homogeneous. We restrict ourselves with
nonrelativistic approximation and neglect the spin of electron. From
the very beginning we take measures to simplify the numerical
computations and to achieve high enough order of perturbation theory.
For this aim we consider $\gamma$ as a fixed parameter, replacing the
double expansion in external fields by the single-variable series
$$
\psi=\sum_{k=0}^\infty {\cal E}^k |k\rangle \ , \hphantom{12345}
E=\sum_{j=0}^\infty E_{2j}^\perp{\cal E}^{2j} \ ,
\eqno(1)
$$
where the wave function corrections $|k\rangle$ and
hyper-polarizabilities $E_k^\perp$ depend on $\gamma$. We introduce
also circular coordinates
$$
x_\pm = x\pm iy \ ,
$$
then all further relations will have real coefficients. In these coordinates
the hamiltonian of our problem is:
$$
\begin{tabular}{ll}
$\hat{H}=\hat{H}_0 + {\cal E}\hat{H}_1 + {\cal E}^2\hat{H}_2 \ ;$
&
\phantom{a}$\displaystyle\hat{H}_0=-\frac{1}{2}\Delta - \frac{1}{r} \ ;$ \\
\multicolumn{2}{l}{
$\displaystyle\hat{H}_1=x+\frac{\gamma}{2}\hat{L}_z=\frac{1}{2}(x_+ + x_-)
 +\frac{\gamma}{2}\left(x_+\frac{\partial}{\partial x_+}
-x_-\frac{\partial}{\partial x_-}\right) \ ;
\phantom{\frac{A^{A^{A^A}}}{A}}$} \\
$\displaystyle
\hat{H}_2=\frac{\gamma^2}{8}(x^2+y^2)=\frac{\gamma^2}{8}x_+x_- \ .$ & \\
\end{tabular}
\eqno(2)
$$
The wave function correction of the order $k$ satisfies the
differential equation
$$
(\hat H_0 -E_0)|k\rangle =-\hat H_1|k-1\rangle -\hat H_2|k-2\rangle
+ \sum \limits _{j=1}^{[k/2]}E_{2j}^\perp|k-2j\rangle \ .
\eqno(3)
$$
Just as in other problems where the moment method was
used~\cite{a,gkw} it is not difficult to transform equation (3) into
algebraic relation between moments of the order $k$
$$
P_{\sigma\alpha\beta}^k =\langle 0|r^{\sigma-\alpha-\beta}
x_+^{\alpha}x_-^{\beta}|k\rangle \ .
\eqno(4)
$$
A recurrence relation results
$$
\frac12(\sigma-\alpha-\beta)(\sigma+\alpha+\beta+1)
P_{\sigma-2,\alpha\beta}^k+2\alpha\beta
 P_{\sigma-2,\alpha-1,\beta-1}^k-\sigma P_{\sigma-1,\alpha\beta}^k
 =R_{\sigma\alpha\beta}^k  \ ,
\eqno(5)
$$
where
$$
R_{\sigma\alpha\beta}^k\equiv\frac12[P_{\sigma+1,\alpha+1,\beta}^{k-1}
+P_{\sigma+1,\alpha,\beta+1}^{k-1}
+\gamma(\alpha-\beta)P_{\sigma\alpha\beta}^{k-1}]
+\frac{\gamma^2}{8}P_{\sigma+2,\alpha+1,\beta+1}^{k-2}
$$
$$
-\sum \limits_{j=1}^{[k/2]}E_{2j}^\perp P_{\sigma\alpha\beta}^{k-2j} \ .
$$
The right-hand side of eq.~(5) and hyper-polarizability
$E_k^\perp$ depend only on the moments of preceding orders. As usual
in the moment method~\cite{a}, the orthogonality condition is accepted
$$
\langle 0|k\rangle=\delta_{0,k} \qquad\longrightarrow\qquad
P_{0,0,0}^k=\delta_{0,k} \ .
\eqno(6)
$$
An expression for hyper-polarizability arises from eq.~(5) at
$\sigma=\alpha=\beta=0$ and even $k$
$$
E_k^\perp=\frac12(P_{1,1,0}^{k-1}+P_{1,0,1}^{k-1})
+\frac{\gamma^2}{8}P_{2,1,1}^{k-2} \ .
\eqno(7)
$$
The closed system of recurrence relations (5) -- (7)
allows to achieve, at least in principle, an arbitrary high order of
perturbation theory. The sequence of operations is similar (also
somewhat simpler) to that, used in the work~\cite{gkw} to compute
Zeeman's shift of a non-degenerate state. At every order $k$ only
moments $P^k_{\sigma\alpha\beta}$ from the sector $\sigma\ge\alpha+\beta-1$,
$\alpha\ge0$, $\beta\ge0$ are necessary. They are evaluated by
successively increasing of $\sigma$, $\alpha$ and $\beta$ values with the
help of eq.~(5).

     We have obtained
hyper-polarizabilities in perpendicular fields up to 80th order,
see table 1. This order is large enough to compare the dependence of
these coefficients on $\gamma$, see fig.~1, with the predictions,
following from quasiclassical considerations. One can see from
fig.~1, that the function $f_k(\gamma)\equiv\ln(|E_k^\perp|/k!)$
has two features. It has a minimum at $\gamma\approx3.4$ and a
sequence of singular points to the right of this value. Besides,
the function $E_k^\perp(\gamma)$ changes it's sign at every singular point
of $f_k(\gamma)$.

     As follows from table 1, at not very large $\gamma$ values all
$E_{2j}^{\perp}$ coefficients have negative sign, as in the case of
Stark effect. In intermediate region of $\gamma$ values the sequence
of $E_{2j}^{\perp}$ signs is irregular and for sufficiently large
$\gamma$'s the series has normal Zeeman's sequence of signs
$(-1)^{j+1}$.

\begin{center}
\bf
3. High orders asymptotic
\end{center}

   As is well known~\cite{bw}, a dispersion relation connects
asymptotic of high orders coefficients $E_k^\perp$ with the
ionization probability of the atom i.e. with the penetrability of
the potential barrier. This relation arises as a consequence of the fact,
that the energy eigenvalue $E=E_0({\cal E}^2)-\frac i2\Gamma({\cal E}^2)$
has essential singularity at ${\cal E}^2=0$ and a cut along ${\cal
E}^2 > 0$ semiaxis. (And similarly $E({\cal H}^2)$ has essential
singularity at ${\cal H}=0$ and a cut ${\cal H}^2 < 0$.)

   To evaluate the ionization probability $\Gamma$ the imaginary time
method was previously developed~\cite{ppt, pp, pkp}. The leading term
of the asymptotic $\tilde E_k^\perp$ of $E_k^\perp$ coefficients at
$k \to \infty$ is determined by the classical subbarrier path with
extremal value of the abbreviated action. Time takes complex values
during this subbarier motion. There are two kinds of complex
classical trajectories. Like in the Stark effect case, the ionization
may be caused by electric field, at stabilizing influence of the
magnetic field. The path of this kind creates the asymptotic
$$
\tilde E_k^\perp(\gamma) \sim k!\:a^k(\gamma) \ , \quad k \mbox{ is even} \ ,
\eqno(8)
$$
at not very large magnetic field, for $\gamma$ below some critical
value $\gamma_c$. According to~\cite{ps2} $\gamma_c=3{.}54$ for
perpendicular external fields. And it is possible to cross the
barrier also at ${\cal H}^2 < 0$, like in the Zeeman effect problem.
Subbarrier trajectories of this kind are responsible for the form of
$\tilde E_k^\perp(\gamma)$ in the opposite case $\gamma > \gamma_c$. This
change of asymptotic explains the origin of the left minimum in fig.~1.

   Having in mind to get estimate for the function $a(\gamma)$,
entering $\tilde E_k^\perp$, we apply the results of~\cite{ps2,pkm}
and write here some necessary expressions for the special case of
perpendicular external fields. More general considerations related to
arbitrary $\vec{\cal E}$ and $\vec{\cal H}$ mutual orientation are
contained in the work~\cite{ps2}.

   The time of subbarrier motion satisfies the equation~\cite{pkm}:
$$
\tau^2-(\tau cth\tau-1)^2=\gamma^2
\eqno(9)
$$
which has a set of solutions $\tau_n=in\pi+\tau'_n$.
The minimal value of the imaginary part of
the subbarrier action is provided by $\tau_0$ for $\gamma < \gamma_c$
and by a pair of solutions $\tau_{\pm1}$ for $\gamma >\gamma_c$.
In the region $\gamma<\gamma_c$ the energy half-width is
$$
\Gamma({\cal E}^2)=\frac{B(\gamma)}{\cal E}
\exp{\left[-\frac{2g(\gamma)}{3{\cal E}}\right]} \ ,
\quad g(\gamma)=\frac{3\tau}{2\gamma^3}\left(\gamma^2-\sqrt{\tau^2-\gamma^2}
\right) \ .
\eqno(10)
$$
The dispersion relation in ${\cal E}^2$ then leads to
$$
\tilde{E}_{2j}^\perp = -\frac1{2\pi}\int\limits_0^\infty
\frac{\Gamma(z)dz}{z^{j+1}}
\sim (2j)!\:a^{2j},
\eqno(11)
$$
where
$$
a(\gamma) = \frac3{2g(\gamma)} \ .
\eqno(12)
$$
The last equality is valid also in the region $\gamma > \gamma_c$,
where $g(\gamma)$ and $a(\gamma)$ are complex functions.
At $\gamma<\gamma_c$ the resulting approximate expressions for $a(\gamma)$
are
$$
a(\gamma)=\frac{3}{2}\left(1-\frac{1}{30}\gamma^2 -\frac{71}{2100}
\gamma^4 + \cdots \right) \ , \qquad \gamma \ll 1 \ ;
\eqno(13)
$$
$$
a(\gamma) \simeq \frac{4\gamma^3}{(\gamma^2-1)^2(1-2e^{-\gamma^2-1})} \ ,
\qquad \gamma\gg1 \ .
\eqno(14)
$$
And in the region $\gamma > \gamma_c$ another representation works
$$
|a(\gamma)|=\frac{\gamma}{\pi}\left[1-\frac 2{\gamma^2}+\left(
 \frac{8\pi^2}{3}+3 \right)\frac 1{\gamma^4} + \cdots \right]\:.
\eqno(15)
$$
On the other hand in the limit of large $k$ a simple relation appropriate for
numerical evaluation holds:
$$
ln\:|a(\gamma)| = \frac{d}{dk}\:ln\frac{|E_k|}{k!} \ .
\eqno(16)
$$
Evaluating $a(\gamma)$ above $\gamma_c$,
we used smoothed function $E_k^{\perp}(\gamma)$, with the nodes
vicinities excluded. A comparison of this way numerically obtained
function $a(\gamma)$ with expressions (13) -- (15) is presented in
fig.~2.

   Now we turn our attention to the region $\gamma > \gamma_c$.
Two solutions of eq.~(9) $\tau_1$ and $\tau_{-1}$ lead to complex
conjugate values of $g(\gamma)$. Substituting approximate $\tau_1$
value into second expression (10), it is possible to get the
phase of the function $a(\gamma)$:
$$
arg\:(a)=-arg\:(g)=-\frac{\pi}2 +\alpha(\gamma),\qquad
\alpha(\gamma)=\frac2{\gamma} - \frac{\pi^2+2}{3\gamma^3} + O(1/\gamma^5) \ .
$$
Finally the sign-alternating asymptotic arises:
$$
\tilde E_{2j}^\perp =2\:|B(\gamma)|\:(2j)!\:|a|^{2j+1}
\cos{\left[(2j+1)\left(-\frac \pi 2 +\alpha(\gamma)\right) +
\beta(\gamma)\right]}
$$
$$
\sim
(-1)^j\:(2j)!\:|a|^{2j+1}\sin{[(2j+1)\alpha(\gamma)+\beta(\gamma)]} \ , \qquad
j\gg1 \ .
\eqno(17)
$$
Here $\beta(\gamma)=arg\:(B)$ is the phase of the preexponential
factor. Its relative contribution to the total phase falls like
$1/j$.

   When the order of perturbation $2j$ is fixed and $\gamma$
increases, expression (17) changes its sign at every point where the
argument of the sinus turns to zero.
This could explain the singular points in fig.~1 in the language of
asymptotic. But rather lengthy calculations are required
to establish detailed quantitative correspondence between asymptotic (17)
and exact $E^{\perp}_{2j}$ coefficients, including nodes vicinities.
Simple approximate expression for $\alpha(\gamma)$ is not enough for this aim.

\begin{center}
\bf
4. Discussion
\end{center}

For the general case of the ground state energy expansion in powers of crossed
external fields, the term of the forth power was known long enough~\cite{lvhk}.

$$
E=-\frac{1}{2}+\sum^{\infty}_{j=1}E^{(2j)} \ ; \qquad
\ E^{(2)}=-\frac{4}{9}\vec{\cal E}^2+\frac{1}{4}\vec{\cal H}^2 \ ;
\eqno(18)
$$
$$
E^{(4)}=-\frac{3555}{64}\vec{\cal E}^4
+\frac{159}{32}\vec{\cal E}^2\vec{\cal H}^2
+\frac{10}{3}[\vec{\cal H}\vec{\cal E}]^2-\frac{53}{192}\vec{\cal H}^4 \ .
\eqno(19)
$$

The value of $E^{(4)}$ is confirmed for perpendicular fields by the
work~\cite{mmo}
and for parallel fields -- by~\cite{mmo,ps,jsf}.
The $E^{\perp}_4$ coefficient,
computed by means of recurrence relations (5) -- (7) exactly agree with
(19). But we have noticed numerical difference between our coefficient
$E^{\perp}_6$ and corresponding quantity from the work~\cite{mmo}. Therefore
the sixth order of perturbation theory was analyzed in details.

     The magneto-electric susceptibilities, i.e. coefficients of the double
series in powers of external fields, can be easily obtained from
hyper-polarizabilities $E^{\perp}_k(\gamma)$. Thus, in the sixth order, taking
into account that Stark's and Zeeman's coefficients are fixed, it is enough
to choose four different $\gamma$ values and to solve the system of four
linear equations. The following representation results
$$
E^{\perp}_6=-\frac{1}{512}\left(2512779-521353\gamma^2
+\frac{953869}{27}\gamma^4-\frac{5581}{9}\gamma^6\right)
$$
$$
\equiv\sum^3_{j=0}\gamma^{\perp}_{6-2j,2j}({\cal H}/{\cal E})^{2j} \ .
\eqno(20)
$$
(The last identity introduces notation of~\cite{mmo}.) Using linear relation
between expansions (1) and (18) and the known magneto-electric
susceptibilities in parallel fields~\cite{jsf}, it is easy to obtain another
term of series (18):
$$
E^{(6)}=-\frac{2512779}{512}\vec{\cal E}^6
+\frac{254955}{512}\vec{\cal E}^4\vec{\cal H}^2
+\frac{133199}{256}\vec{\cal E}^2[\vec{\cal H}\vec{\cal E}]^2
$$
$$
-\frac{49195}{1536}\vec{\cal E}^2\vec{\cal H}^4
-\frac{255557}{6912}\vec{\cal H}^2[\vec{\cal H}\vec{\cal E}]^2
+\frac{5581}{4608}\vec{\cal H}^6 \ .
\eqno(21)
$$
Some next terms of series (18) can be obtained in the same way. Expressions
(20) and (21) are convenient to check term by term the sixth order correction.
As follows from~\cite{mmo}
$$
\gamma^{\perp\cite{mmo}}_{24}=-\frac{1610197}{27648} \quad \mbox{and}
\quad \gamma^{\perp\cite{mmo}}_{42}=\frac{2417015}{1536} \ ,
\eqno(22)
$$
while the results of our computation are
$$
\gamma^{\perp}_{24}=-\frac{953869}{13824} \quad \mbox{and}
\quad \gamma^{\perp}_{42}=\frac{521353}{512} \ .
\eqno(23)
$$
All other corresponding coefficients of~\cite{mmo} and of present work
coincide. We carried out additional independent calculation by means of
the method from the work~\cite{lvhk} and get
$$
\gamma^{\perp\cite{lvhk}}_{24}=-\frac{953869}{13824} \ ,
\eqno(24)
$$
see Appendix. Note, that~\cite{lvhk} contains complete correction of the sixth
power in external fields for the case of parallel fields and only a part of
it for the case of perpendicular fields. These "celebrated" sixth order
terms result as a by-product of forth-order calculations in the
work~\cite{lvhk}.
The agreement between high-order hyper-polarizabilities $E_k^\perp$ and
their asymptotic $\tilde{E_k^\perp}$ presents additional confirmation of
correctness of recurrence relations (5) -- (7).

\begin{center}
\bf
5. Concluding remarks
\end{center}

     The considered above problem demonstrates once more the high
efficiency and convenience of the moment method. The obtained
recurrence relations have allowed to advance up to 80th order
of perturbation theory. Besides the unusual "oscillations" of
hyper-polarizabilities as a function of the ratio of external
fields were noticed. The high orders asymptotical behavior was
analyzed as well. Basic parameters of this asymptotic exactly agree
with those, previously obtained on the ground of quasiclassical
approximation with the help of imaginary time method.

\begin{center}
\bf
Acknowledgments
\end{center}

     The authors would like to express the deep gratitude to professor
V.~S.~Popov and professor A.~E.~Kudryavtsev for valuable comments and
numerous helpful discussions. We are also grateful to professor
F.~M.~Fernandez for drawing our attention to works~\cite{fc,fc2,fm}.

     This work was supported in part by the Russian Foundation for Basic
Research under Grant No.~98-02-17007 (V.~M.~Weinberg). The work of
V.~A.~Gani was supported by the INTAS Grant 96-0457 within the research
program of the International Center for Fundamental Physics in Moscow.

\newpage

\begin{center}
{\bf Table 1.}
Hyper-polarizabilities $E_k^{\perp}$ of the hydrogen ground state.
\end{center}

\begin{tabular}{|r|l|l|l|l|}
\hline
$k$ &
\multicolumn{1}{|c|}{$\gamma=2.0$} &
\multicolumn{1}{|c|}{$\gamma=3.0$} &
\multicolumn{1}{|c|}{$\gamma=6.0$} &
\multicolumn{1}{|c|}{$\gamma=70.0$} \\
\hline
 2 & $-1.2500000$               & $+0.0000000$               & $+6.7500000$               & $+1222.7500$ \\
 4 & $-26.755208$               & $-3.1875000$               & $-114.42188$               & $-6587135.8$ \\
 6 & $-1861.2023$               & $-449.50781$               & $-1167.7324$               & $+1.4083939\times10^{11}$ \\
 8 & $-231011.83$               & $-39518.994$               & $+3563855.9$               & $-5.5211341\times10^{15}$ \\
10 & $-4.3046334\times10^{7}$   & $-4415104.3$               & $-1.9148046\times10^{9}$   & $+3.2420587\times10^{20}$ \\
12 & $-1.1108858\times10^{10}$  & $-7.8928562\times10^{8}$   & $+8.7798001\times10^{11}$  & $-2.6154136\times10^{25}$ \\
14 & $-3.7903062\times10^{12}$  & $-1.9681752\times10^{11}$  & $-2.7563534\times10^{14}$  & $+2.7647695\times10^{30}$ \\
16 & $-1.6565997\times10^{15}$  & $-6.0102169\times10^{13}$  & $-1.4386041\times10^{17}$  & $-3.7128281\times10^{35}$ \\
18 & $-9.0515867\times10^{17}$  & $-2.2599584\times10^{16}$  & $+5.1094372\times10^{20}$  & $+6.1877660\times10^{40}$ \\
20 & $-6.0598915\times10^{20}$  & $-1.0569584\times10^{19}$  & $-8.5724488\times10^{23}$  & $-1.2555439\times10^{46}$ \\
22 & $-4.8865029\times10^{23}$  & $-5.9768835\times10^{21}$  & $+1.0371317\times10^{27}$  & $+3.0513954\times10^{51}$ \\
24 & $-4.6763388\times10^{26}$  & $-3.9866393\times10^{24}$  & $-3.8968989\times10^{29}$  & $-8.7572953\times10^{56}$ \\
26 & $-5.2434742\times10^{29}$  & $-3.1103268\times10^{27}$  & $-3.2330568\times10^{33}$  & $+2.9313099\times10^{62}$ \\
28 & $-6.8121442\times10^{32}$  & $-2.8159706\times10^{30}$  & $+1.5576520\times10^{37}$  & $-1.1320123\times10^{68}$ \\
30 & $-1.0154266\times10^{36}$  & $-2.9246470\times10^{33}$  & $-4.7085047\times10^{40}$  & $+4.9954972\times10^{73}$ \\
40 & $-4.4829424\times10^{52}$  & $-2.1028668\times10^{49}$  & $-6.2371218\times10^{58}$  & $-4.7347888\times10^{102}$ \\
50 & $-2.3374671\times10^{70}$  & $-1.7764719\times10^{66}$  & $-8.3813757\times10^{77}$  & $+4.7994225\times10^{132}$ \\
60 & $-8.8335861\times10^{88}$  & $-1.0843471\times10^{84}$  & $-5.8480016\times10^{97}$  & $-3.1030447\times10^{163}$ \\
70 & $-1.7441216\times10^{108}$ & $-3.4513657\times10^{102}$ & $+1.9127476\times10^{117}$ & $+8.3761299\times10^{194}$ \\
80 & $-1.4229765\times10^{128}$ & $-4.5336207\times10^{121}$ & $+1.3366049\times10^{140}$ & $-4.3789967\times10^{226}$ \\
\hline
\end{tabular}
ÿ
\begin{center}
\bf
Appendix
\end{center}

     Extending the described in the work~\cite{lvhk} calculations we obtained,
by the same method, the ground state energy correction, which is proportional
to $Q\equiv{\cal H}^2[\vec{\cal H}\vec{\cal E}]^2$.
The perturbation in~\cite{lvhk} includes Stark's term
$V_s=\vec{\cal E}\vec{r}$, paramagnetic
$V_p=\frac{1}{2}\vec{\cal H}\hat{\vec{L}}$ and diamagnetic
$V_D=\frac{1}{8}[\vec{\cal H}\vec{r}]^2$ terms. The entire perturbation is
inhomogeneous, therefore terms of the sixth power in external fields are
presented in corrections of the fourth, fifth and sixth orders in $V$.
$$
\varepsilon^{(4)}=-\frac{151347}{2047}Q+... \ ,
\eqno(A1)
$$
$$
\varepsilon^{(5)}=\langle2|(V-\varepsilon^{(1)})|2\rangle
-2\varepsilon^{(2)}\langle2|1\rangle-\varepsilon^{(3)}\langle1|1\rangle \ ,
\eqno(A2)
$$
$$
\varepsilon^{(6)}=\langle3|(V-\varepsilon^{(1)})|2\rangle
-\varepsilon^{(2)}(\langle2|2\rangle+\langle1|3\rangle)
-2\varepsilon^{(3)}\langle2|1\rangle-\varepsilon^{(4)}\langle1|1\rangle \ .
\eqno(A3)
$$
In the following an abbreviated mnemonic notation will be used, reflecting
the origin of each term and the powers of entering this term external fields.
This notation helps to omit all not essential terms. Operation signs are
encircled in the abbreviated notation. In the first order in $V$
$$
|1\rangle\equiv\{a_1(r)(\vec{\cal E}\vec{r})+
\frac{1}{4}(a_2(r){\cal H}^2+a_3(r)[\vec{\cal H}\vec{\cal E}]^2)\}|0\rangle
\rav\{V_{\cal E}\pls V_D\}|0\rangle \ .
\eqno(A4)
$$
The next correction $|2\rangle$ contains
$$
V_P V_{\cal E}|0\rangle
\rav\frac{i}{2}b_1(r)([\vec{\cal H}\vec{\cal E}]\vec{r})|0\rangle \ ,
$$
$$
V_{\cal E} V_D|0\rangle
\rav\frac{1}{4}\{(\vec{\cal E}\vec{r})(b_4(r)[\vec{\cal H}\vec{r}]^2
+b_5(r){\cal H}^2)+b_6(r){\cal H}^2([\vec{\cal H}\vec{\cal E}]\vec{r})\}
|0\rangle \ .
\eqno(A5)
$$
The polynomials $a_i(r)$ and $b_i(r)$ are given in the article~\cite{lvhk}.
Abbreviated notation allows to verify that in each of the right-hand sides
of Eqs. (A2) and (A3) only the first matrix element yields contribution
$\sim Q$.
$$
\varepsilon^{(5)}=A+B_1+B_2+C+... \ ,
$$
dots stand for all omitted terms.
$$
\begin{tabular}{ll}
\multicolumn{2}{l}{
$A\rav\langle 0|(V_P V_{\cal E})V_D(V_P V_{\cal E})|0\rangle \ ,$}\\
\multicolumn{2}{l}{
$B_1\rav\langle 0|(V_{\cal E}V_D)V_P(V_P V_{\cal E})|0\rangle \ ,$}\\
$B_2\rav\langle 0|(V_P V_{\cal E})V_P(V_{\cal E}V_D)|0\rangle \ ,$ &
$B_2=B_1 \ ,$ \\
\multicolumn{2}{l}{
$C\rav-\varepsilon^{(1)}\langle 0|(V_P V_{\cal E})(V_P V_{\cal E})|0\rangle
\ .$}
\end{tabular}
\eqno(A6)
$$
Corresponding explicit expressions are
$$
\begin{tabular}{l}
$\displaystyle
A=\frac{1}{32}\langle 0|b_1^2(r)([\vec{\cal H}\vec{\cal E}]\vec{r})^2
[\vec{\cal H}\vec{r}]^2|0\rangle=\frac{571}{48}Q \ ,$ \\
$\displaystyle
B_1=\ \frac{1}{16}\langle 0|b_1(r)\{b_4(r)[\vec{\cal H}\vec{r}]^2
+(b_5(r)+b_6(r)){\cal H}^2\}([\vec{\cal H}\vec{\cal E}]\vec{r})^2|0\rangle$ \\
$\displaystyle
\phantom{1111111111}=\frac{299623}{18432}Q=B_2 \ ,$ \\
$\displaystyle
C=-\frac{1}{16}{\cal H}^2\langle 0|b_1^2(r)
([\vec{\cal H}\vec{\cal E}]\vec{r})^2
|0\rangle=-\frac{9673}{4608}Q \ .$ \\
\end{tabular}
\eqno(A7)
$$
Only one term of the third correction to the wave function is essential --
that of the lowest power in external fields:
$$
|3\rangle =\frac{1}{4}c_1(r)([\vec{\cal H}[\vec{\cal H}\vec{\cal E}]]\vec{r})
|0\rangle+...\rav(V_P V_{\cal E}\pls...)|0\rangle \ .
\eqno(A8)
$$
The differential equation for $c_1(r)$~\cite{lvhk} is satisfied by
the polynomial:
$$
c_1(r)=-\frac{1}{144}(450+225r+62r^2+6r^3) \ .
\eqno(A9)
$$
As a consequence we get
$$
\varepsilon^{(6)}\rav\langle 0|(V_P^2V_{\cal E})V_P(V_P V_{\cal E})
|0\rangle+...
\eqno(A10)
$$
and
$$
\varepsilon^{(6)}=-\frac{1}{16}\langle 0|b_1(r)c_1(r)
([\vec{\cal H}[\vec{\cal H}\vec{\cal E}]]\vec{r})^2|0\rangle+... \ .
\eqno(A10')
$$
The total energy correction of the desired form is
$$
\Delta E=-\frac{255557}{6912}{\cal H}^2[\vec{\cal H}\vec{\cal E}]^2+... \ .
\eqno(A11)
$$
One should not forget also the "isotropic" contribution to the energy
correction, originating from $\varepsilon^{(4)}$:
$$
-\frac{49195}{1536}{\cal H}^4{\cal E}^2 \ .
$$

\begin{center}
\bf
Figure captions
\end{center}

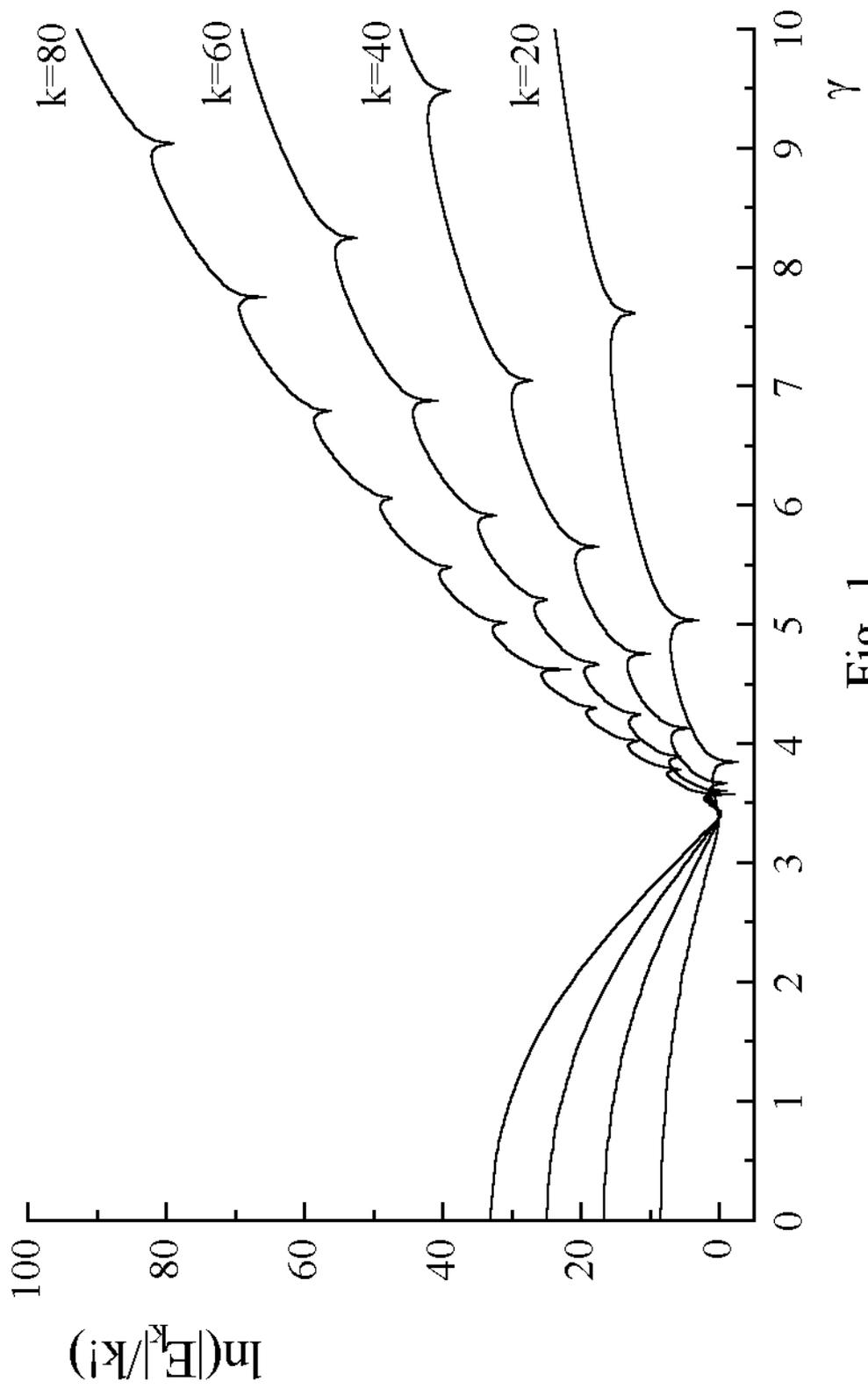
{\bf Fig.~1.} Functions
              $\displaystyle f_k(\gamma)=\ln{\frac{|E_k^\perp|}{k!}}$
              resulting from the recurrently computed hyper-polarizabilities.

\bigskip

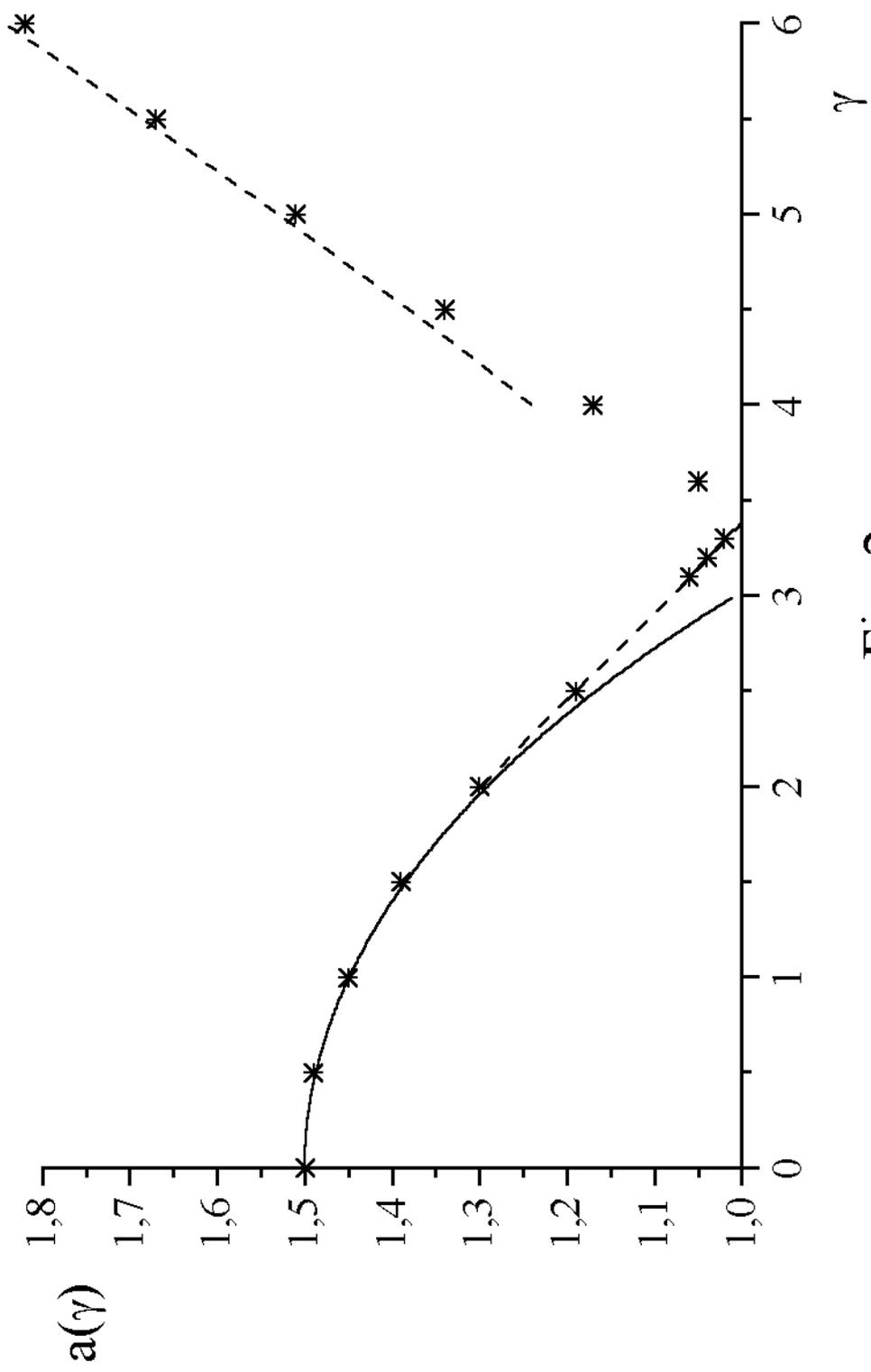
{\bf Fig.~2.} Parameter $a(\gamma)$ of the perturbation series asymptotic.\\
              The solid line follows from the quasiclassical estimate at
              $\gamma\ll1$, see eq.~(13);
              the same estimate for $\gamma\gg1$ is presented by dashed lines,
              see eqs.~(14) and (15).
              Numerically obtained values are denoted by stars.
ÿ

\end{document}